\documentclass[12pt]{article}
\usepackage{amsfonts,amssymb,amsmath,graphicx}
\begin{document}

\begin{center}
\begin{Large}
{The theory of the reentrant effect in susceptibility of cylindrical
mesoscopic samples}
\end{Large}
\end{center}
\bigskip

\begin{center}
{G.A. Gogadze\\
{\it B. Verkin Institute for Low Temperature Physics and Engineering}\\
{\it of the National Academy of Sciences of Ukraine}\\
47, Lenin Av., 61103 Kharkov Ukraine\\
e-mail: {\it gogadze@ilt.kharkov.ua}}
\end{center}
\bigskip

\abstract{
A theory has been developed to explain the anomalous behavior of the
magnetic susceptibility of a normal metal-superconductor ($NS$) structure
in weak magnetic fields at millikelvin temperatures. The effect was discovered
experimentally by A.C. Mota et al \cite{10}. In cylindrical superconducting
samples covered with a thin normal pure metal layer, the susceptibility exhibited
a reentrant effect: it started to increase unexpectedly when the temperature
lowered below $100$ mK. The effect was observed in mesoscopic $NS$ structures
when the $N$ and $S$ metals were in good electric contact. The theory proposed
is essentially based on the properties of the Andreev levels in the normal metal.
When the magnetic field (or temperature) changes, each of the Andreev levels coincides
from time to time with the chemical potential of the metal. As a result,
the state of the $NS$ structure experiences strong degeneracy, and the quasiparticle
density of states exhibits resonance spikes. This generates a large paramagnetic
contribution to the susceptibility, which adds up to the diamagnetic contribution
thus leading to the reentrant effect. The explanation proposed was obtained within
the model of free electrons. The theory provides a good description for experimental
results \cite{10}.
}

\begin{large}
\section{Introduction}
\*
\par Mesoscopic systems \cite{1}--\cite{3} can exhibit surprising properties at
comparatively low temperatures. For pure normal metals there is a length scale
$\xi_N = \hbar V_F/k_B T$ ($V_F$ is the Fermi velocity, $T$ is the temperature,
$k_B$ is the Boltzmann constant) which has the meaning of a coherence length
in a system with a disturbed long-range order. When this length is comparable with
the characteristic dimensions of the system, the interference effects can come into
play. Theoretically this was first demonstrated by Kulik \cite{4} for a thin-wall
normal pure-metal cylinder in the vector potential field. It appears that the magnetic
moment of such a system is an oscillating function of the magnetic flux through the
cross-section of the cylinder, the oscillation period being equal to the flux quantum
of the normal metal $hc/e$. The effect is generated by quantization of the electron
motion and due to the sensitivity of the states of the system to the vector potential
field (Aharonov--Bohm effect \cite{5}). Bogachek and this author showed the existence
of oscillating component with the period $hc/e$ in the magnetic moment
of a singly connected normal cylinder in a weak magnetic field. Oscillations with
this period are produced by the magnetic surface levels of the cylindrical sample
in a weak magnetic field \cite{6}. The effect of flux quantization in a normal
singly connected cylindrical conductor was first detected experimentally in 1976
by Brandt et al. when they were investigating the longitudinal magnetoresistance
in pure $Bi$ single crystals \cite{7},\cite{8}. This was actually the first observation
of the interference effect of flux quantization in nonsuperconducting condensed
matter.

Recent advanced technologies of preparation of pure samples have enabled investigation
of the coherent properties of mesoscopic structures taking proper account of the
proximity effect \cite{9}. The samples were superconducting $Nb$ wires with a radius
$R$ of tens of $\mu m$ coated with a thin layer $d$ of high-purity $Cu$ or $Ag$.
The metals were in good contact and the electron mean free path exceeded the typical
scale $\xi_N$. The magnetic susceptibilities of copper and silver were measured. The
breakdown field $H_b$, the supercooled field $H_{sc}$ and the superheated field
$H_{sh}$ were estimated as functions of temperature and normal metal thickness.
While continuing their experiments on these samples, Mota and co-workers \cite{10}
detected a surprising behavior of the magnetic susceptibility of a cylindrical
$NS$ structure ($N$ and $S$ are for the normal metal and the superconductor, respectively)
at very low temperatures ($T < 100 mK$) in the external magnetic field parallel
to the $NS$ boundary.

Most intriguingly, a decrease in the sample temperature below a certain point $T_r$
(at a fixed field) produced a reentrant effect: the decreasing magnetic susceptibility
of the structure unexpectedly started growing. A similar behavior was observed
with the isothermal reentrant effect in a field decreasing to a certain value
$H_r$ below which the susceptibility started to grow sharply. It is emphasized
in Ref.\cite{11} that the detected magnetic response of the $NS$ structure is similar
to the properties of the persistent currents in mesoscopic normal rings. It is assumed
\cite{9} -- \cite{12} that the reentrant effect reflects the behavior of the total
susceptibility $\chi$ of the $NS$ structure: the paramagnetic contribution is
superimposed on the Meissner effect-related diamagnetic contribution and nearly
compensates it. Anomalous behavior of the susceptibility has also been observed
in $AgTa$, $CuNb$ and $AuNb$ structures \cite{11}, \cite{13}.

The possibility of the paramagnetic contribution to the susceptibility of the $NS$
structure needs further clarification. The $NS$ structure in question is essentially a
combination of two subsystems capable of electron exchange, which corresponds to the
establishment of equilibrium in a large canonical ensemble (with fixed chemical
potential). Assume that these systems are initially isolated with a thick dielectric
layer. It is known that the superconductor response to the applied magnetic field
generates superfluid screening current near the cylinder surface (Meissner effect).
How does the normal mesoscopic layer respond to the weak magnetic
field? Kulik \cite{4} shows (see above) that in a weak magnetic field the magnetic
moment of a thin-wall normal cylinder oscillates with the flux. The magnetic moment
oscillations are equivalent to the existence of persistent current. Since the energies
of the individual states and< hence, the total energy are dependent on the flux,
the average current is nonzero. The current state corresponds to the minimum free
energy, therefore the inclusion of weak dissipation would not lead to the decay
of the current state. When the $N$ and $S$ metals are isolated, the quantum states
of the quasiparticles in the $N$-metal are formed at the expense of specular reflection
of the electrons from the dielectric boundaries. The amplitude of the magnetic
moment oscillations in the $N$ layer is small, which is determined by the smallness
of the parameter $1/k_F R$ in the problem and by the paramagnetic character of the
persistent current \cite{4},\cite{6} (when the magnetic field tends to zero,
the magnetic susceptibility is positive). Thus, in the absence of the proximity effect,
the total susceptibility of the $NS$ structure is only governed by the diamagnetic
contribution of the $S$-layer (the paramagnetic contribution is very small).

When the proximity effect is present in the $NS$ structure, we assume that the
probability of the electron transit from the superconductor to the $N$ metal is close
to unity. This significantly affects the properties of the $NS$ structure. The diamagnetic
response of the superconductor persists but new properties appear, that are brought
about by the proximity effect. Now two kinds of electron reflection are observed
in the normal film -- a specular reflection from one boundary and the Andreev reflection
from other. Along with the trajectories closed around the cylinder circle, new trajectories
appear in a weak field, which "screen" the normal metal. The new trajectories of
"particles" and "holes" confine the quantization area of the triangle whose base
is a part of the $NS$ boundary between the points of at which the quasiparticle
collides with this boundary. This area is maximum for the trajectories touching the superconductor.
It is shown below that at certain values of the flux through the triangle area, the
electron density of states experiences flux-dependent resonance spikes. Thus, in the
presence of the proximity effect, the periodic flux-induced oscillations of the
thermodynamic values typical of the normal layer in the $NS$ structure give way to
periodic resonance spikes with a period equal to a superconducting flux quantum
$hc/2e$ \cite{16}. The response of the normal mesoscopic layer to a weak magnetic
field ($H\lesssim 10 Oe$) is
paramagnetic and the susceptibility amplitude is large. The picture, however, changes
when the quantized magnetic flux through the triangle area increases and its value
divided by $hc/2e$ starts to exceed the highest Andreev "subband" number. A phase
transition occurs in a certain field $H_r$. As a result, the $N$ layer is now screened
only by the trajectories of those quasiparticles that do not collide with the
superconducting boundary. Their amplitudes are rather small (see above) against the
large diamagnetic response. We can thus conclude that the resonance contribution
to the paramagnetic susceptibility of the $NS$ structure can only appear in comparatively
weak magnetic fields. At this condition the reentrant effect may be generated.
The conclusion correlates well with the experimental observations \cite{9} --
\cite{14}.

The origin of paramagnetic currents in $NS$ structure was discussed in several
theoretical publications. Bruder and Imry \cite{17} analyze the paramagnetic contribution
to susceptibility made by quasiclassical ("glancing") trajectories of quasiparticles
that do not collide with the superconducting boundary. The authors \cite{17} point
to a large paramagnetic effect within their physical model. However, their
ratio between the paramagnetic and diamagnetic contributions is rather low and cannot
account for experimental results \cite{9} -- \cite{14}.

Fauchere, Belzig and Blatter \cite{18} explain the large paramagnetic effect assuming a
pure repulsive electron--electron interaction in noble metals. The proximity
effect in the $N$ metal induces an order parameter whose phase is shifted by $\pi$
from the order parameter $\Delta$ of the superconductor. This generates the paramagnetic
instability of the Andreev states, and the density of states of the $NS$ structure
exhibits a single peak near the zero energy. The theory in \cite{18} essentially rests
on the assumption of the repulsive electron interaction in the $N$ metal. Is the
reentrant effect a result of specific properties of noble metals? or Does it display the
behavior of any normal metal experiencing the proximity effect from the neighboring
superconductor? Only experiment can provide answers to these questions. We just
note that the theories of \cite{17}, \cite{18} do not account for the temperature
and field dependencies of the paramagnetic susceptibility and the nonlinear behavior
$\chi$ of the $NS$ structure. The current theories cannot explain the origin
of the anomalously large paramagnetic reentrant
susceptibility in the region of very low temperatures and weak magnetic fields.

It is worth mentioning the assumption made by Maki and Haas \cite{19} that below
the transition temperature ($\sim 10 mK$) some noble metals ($Cu$, $Ag$, $Au$)
can exhibit $p$-ware superconducting ordering, which may be responsible for the
reentrant effect. This theory does not explain the high paramagnetic reentrant
effect either.

In this paper a theory of the reentrant effect is proposed which is essentially
based on the properties of the quantized levels of the $NS$ structure. Levels with
energies no more than $\Delta$ ($2\Delta$ is the gap of the superconductor)
appear inside the normal metal bounded by the dielectric (vacuum) on one side and
contacting the superconductor on the other side. The number of levels $n_0$ in the
well is finite. Because of the Aharonov-Bohn effect \cite{5}, the spectrum of
the $NS$ structure is a function of the magnetic flux in a weak field. The specific
feature of the quantum levels of the structure is that in a varying field $H$ (or
temperature $T$) each level in the well periodically comes into coincidence with
the chemical potential $\zeta$ of the metal. As a result, the state of the system
suffers strong degeneracy and the density of states of the $NS$ sample experiences
resonance spikes.

It is shown that the phenomenon of resonance appears in a certain interval of weak magnetic
fields at temperatures no higher than a hundred of millikelvins. Resonance is realizable
only in pure mesoscopic $N$ layers under the condition of the Aharonov-Bohm effect.
The resonance produces a large paramagnetic contribution $\chi^p$ to the susceptibility
of the $NS$ structure. When $\chi^p$ is added to the diamagnetic contribution $\chi^d$
produced by the Meissner effect, the total susceptibility displays the features of the
reentrant effect \cite{20}.

\section{Spectrum of quasiparticles of the $NS$ structure}
\*
\par Consider a superconducting cylinder with the radius $R$ which is covered with a
thin layer $d$ of a pure normal metal. The structure is placed in a weak magnetic
field $\vec H (0,0,H)$ oriented along the symmetry axis of the structure.
It is assumed that the field is weak to an extent that the effect of twisting
of quasiparticle trajectories becomes negligible. It actually reduces to the Aharonov-Bohm
effect \cite{5}, i.e. allows for the increment in the phase of the wave function
of the quasiparticle moving along its trajectory in the vector potential field.

We proceed from a simplified model of $NS$ structure in which the order parameter
magnitude changes stepwise at the $NS$ boundary. It is also assumed that the magnetic
field does not penetrate into the superconductor. The coherent properties observed in the
pure normal metal can be attributed to its large "coherence" length $\xi_N$
at very low temperatures.

One can easily distinguish two classes of trajectories inside the normal metal.
One of them includes the trajectories which collide in succession with the
dielectric and $NS$ boundaries. The quasiparticles moving along these trajectories
have energies $\varepsilon <\Delta$ and are localized inside the potential well
bounded by a high dielectric barrier ($\simeq 1 eV$) on one side and by the
superconducting gap $\Delta$ on the other side. On its collisions, the quasiparticle
is reflected specularly from the dielectric and experiences the Andreev scattering
at the $NS$ boundary \cite{15}. We introduce an angle $\alpha$ at which the quasiparticle
hits the dielectric boundary. The angle is counted off the positive direction
of the normal to the boundary (Fig. 1). in this case the first class contains the
trajectories with $\alpha$ varying within $0\lesssim
\alpha \le \alpha_c$ ($\alpha_c$ is the angle at
which the trajectory touches the $NS$ boundary). The other class includes
the trajectories whose spectra are formed by collisions with the dielectric only,
i.e. the trajectories with $\alpha >\alpha_c$.

\begin{figure}
\centering
\includegraphics[height=8cm,width=5cm]{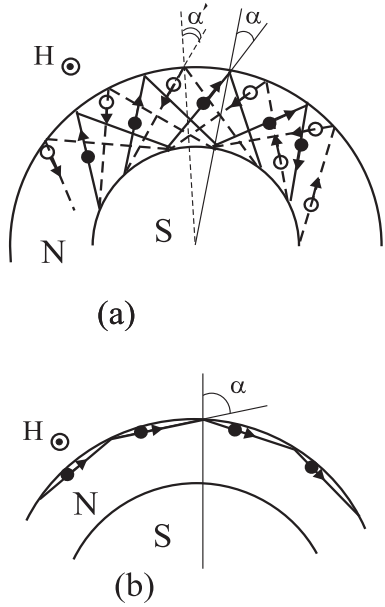}
\caption{Two classes of trajectories in the normal metal of $NS$ structure in 
the magnetic field:
a) trajectories forming the Andreev levels;
b) trajectories colliding only with the dielectric boundary.}
\end{figure}

The two groups of trajectories produce significantly different spectra of quasiparticles.
The distinctions are particularly obvious in the presence of the magnetic field.
The trajectories with $\alpha \lesssim
\alpha_c$ form a spectrum of Andreev levels which contains
a supplement in the form of an integral of the vector potential field. The spectrum
characterizes the magnetic flux through the area of the triangle between the
quasiparticle trajectory and the part of the $NS$ boundary. It is also determines
the magnitude of the screening current produced by "particles" and "holes" in
the $N$ layer. These states are responsible for the reentrant effect. The trajectories
with $\alpha >\alpha_c$ do not collide with the $NS$ boundary. The states induced
by these trajectories are practically similar to the "whispering gallery" type
of states appearing in the cross-section of a solid normal cylinder in a weak magnetic
field \cite{6}, \cite{21}. The size of the caustic of these trajectories is of
the order of the cylinder radius, i.e. they correspond to high magnetic quantum
numbers $m$. The spectrum thus formed carries no information about the parameters
of the superconductor and it is impossible to meet the resonance condition in this
case. These states make a paramagnetic contribution to the thermodynamics of the
$NS$ structure but their amplitude is small ($\sim 1/k_F R$). It is therefore
discarded from further consideration. Our interest will be concentrated on the
trajectories with $\alpha\le \alpha_c$.

The spectrum of quasiparticles of the NS structure can be obtained easily using
the multidimensional quasiclassical method generalized for the case of the Andreev
scattering in the system \cite{16}, \cite{22}. After collision with the $NS$ boundary
the "particle" transforms into a "hole". The "hole" travels practically along the
path of the "particle" but in the reverse direction. In the strict sense, however,
the path of the "hole" is somewhat longer because under the condition of Andreev
elastic scattering the momentum of the "particle" exceeds that of the reflected
"hole". According to the law of conservation of the angular momentum, the angle
$\alpha^\prime$ at which the "hole" comes up to the dielectric boundary and hence the
distance covered by the "hole" are larger. Eventually, the trajectory of the quasiparticle
becomes closed due to its displacement along the perimeter of the $N$ layer.
However, as the quasiparticle energy decreases and approaches the value of the chemical
potential, the difference $\alpha - \alpha^\prime$ starts tending to zero.
Since our further interest is concerned with low-lying Andreev levels, we assume
that the "hole" trajectory is strictly reversible. The distance covered by the
"particle" ("hole") between two boundaries is ${\cal L}_0 \simeq 2d/\cos \alpha$.

According to the multidimensional quasiclassical method \cite{16}, \cite{22},
there are two congruences of "particle" rays -- towards the dielectric ($I$)
and in the opposite direction ($II$). There are also two congruences of "hole" rays
-- towards the $NS$ boundary ($III$) and away from it ($IV$). The covering space
is constructed of four similar $NS$ structures whose edges are joined in accordance
with the law of quasiparticle reflection from a dielectric and a $NS$ boundary.
At the dielectric boundary the congruences $I$ and $II$ are joined. The congruences
$III$ and $IV$ are joined independently. The covering space consists of the outer
("particles") and inner ("holes") toroidal surfaces. Each surface contains only
a part of the single independent integration contour. The path of the "particle"
is $2d$. The "hole" travels the same length whereupon the trajectory of the quasiparticle
closes. The total length of the closed contour along the covering surface of the
$NS$ structure is $4d$.

It is possible to choose two independent integration contours within a tours that
do not contract into a point. One condition of quantization relates the caustic
radius to the magnetic quantum number $m$. We replace it with an angle of incidence
of the quasiparticle on the dielectric boundary. The other condition of quantization
introduces the radial quantum number $n$. Thus, the complete set of quantum numbers
describing the motion of the quasiparticle includes $n$, $\alpha$, $q$, where
$q$ is the quasimomentum component along the symmetry axis of the cylinder.

Assume that the condition $d\ll R$ is obeyed for the $NS$ structure. We can then
neglect the curvature of the cylinder boundary and assume that it is flat. The condition
of quasiclassical quantization can be written as
\begin{equation}
\int\limits_{{\cal L}_0} \left ( \vec p_0 - \frac{\vert e\vert}{c}\vec A\right )
d\vec s- \int\limits_{{\cal L}_0} \left ( \vec p_1 +
\frac{\vert e\vert}{c}\vec A\right ) d\vec s = 2\pi \hbar \left (n+1+
\frac{1}{\pi}\arccos\varepsilon /\Delta\right ),
\label{1}
\end{equation}
where $p_0$ ($p_1$) are the quasimomentum of the "particle" ("hole"), $\varepsilon$
is the "quasiparticle" energy, $\vec A$ is the vector potential $(0,0,Hy)$,
$\vert{\cal L}_0\vert$ is the trajectory length covered by the "particle" ("hole").
The unity in the right-hand side of Eq. (\ref{1}) appears when two collisions of the
quasiparticle with the dielectric boundary are taken into account \cite{22}. The
term $\frac{1}{\pi}\arccos \varepsilon /\Delta$ accounts for the phase delay of the
wave function under the Andreev scattering of quasiparticles \cite{16}. The quasimomentum
$p_0$ and $p_1$ in Eq. (\ref{1}) can be expanded in the parameter $\varepsilon
/\zeta$ retaining the first-order terms and replacing $n+1$ by $n$. As a result,
Eq. (\ref{1}) furnishes the sought for spectrum of the $NS$ structure in a weak magnetic
field (${\cal L}$ is the quasiparticle trajectory):
\begin{equation}
\varepsilon_n (q,\alpha; \Phi) = \frac{\pi \hbar v_{\cal L} (q) \cos\alpha}
{2d} \left ( n+\frac{1}{\pi}\arccos\frac{\varepsilon}{\Delta} -
\frac{{\rm tg}\alpha}{\pi} \Phi\right ).
\label{2}
\end{equation}
Here $v_{\cal L} (q) = \sqrt{p_F^2 -q^2}/m^*$, $p_F$ is the Fermi momentum, $q$
is the quasiparticle momentum component along the cylinder axis, $m^*$ is the
effective mass of the quasiparticle, $\Phi_0 = hc/2e$ is the superconducting flux
quantum. The positive $\alpha$-values refer to "particles" ($n>0$), while the
negative ones are for "holes" ($n<0$).

The last term in Eq.(\ref{2}) has the meaning of "phase"
\begin{equation}
\Phi = \frac{2\pi}{\Phi_0} \int\limits_0^d A(x) dx,
\label{3}
\end{equation}
which is dependent on the vector potential field and varies with the angle $\alpha$
characterizing the trajectory of the quasiparticle.

The spectrum of Eq. (\ref{2}) is similar to Kulik's spectrum \cite{23} for the
current state of an $SNS$ contact. However, Eq. (\ref{2}) includes an angle-dependent
magnetic flux instead of the phase difference of the contacting superconductors.

The value of the "phase" (flux) controls the diamagnetic and paramagnetic
currents in the $NS$ structure. To calculate it, we should know the distribution
of the vector potential field inside the normal metal.

The problem of the Meissner effect in superconductor-normal metal (proximity)
sandwiches was solved by Zaikin \cite{24}. It was shown that the proximity effect
caused the Meissner effect bringing an inhomogeneous distribution of the vector
potential field over the $N$ layer of the structure:
$
A(x) = Hx + \frac{4\pi}{c} j(a)x(d-\frac{x}{2}).
$
For convenience we introduce the notation $a=\int\limits_0^d A(x) dx $.
This expression can be obtained from the Maxwell equation ${\rm rot} \vec H =
\frac{4\pi}{c} \vec j$ with the boundary conditions $A(x=0)=0$ and $\partial_x
A(x=d) =H$. The screening (diamagnetic) current $\vec j$ is a function of $a$, $j(a)
=-j_s \varphi(a/\Phi_0)$, where $j_s$ is the superfluid current and $\varphi (x)$
is function of flux. Thus, we can write down the self-consistent equation for
$a$ \cite{25} -- \cite{26}:
\begin{equation}
a=\frac{Hd^2}{2}+\frac{4\pi}{3c} j(a)d^3.
\label{4}
\end{equation}

The diamagnetic current $\vec j^d (a)$ was calculated in terms of the microscopic
theory as a sum of currents of quasiparticles ("particles" and "holes") for all
quasiclassical trajectories characterized by the angles $\theta$ and $\varphi$
\cite{24}, \cite{26} (below the system of units $k_B = \hbar=c=1$ is used):
\begin{equation}
\begin{split}
& j^d (\Phi, T) = \\
&= -AT\sum_{\omega_n >0} \int\limits_0^{\pi/2} d\theta
\int\limits_0^{\pi/2} d\varphi \frac{\sin^2\theta\cos\varphi\sin[2\Phi{\rm tg}
\theta\cos\varphi]}{\left [\frac{\sqrt{\omega^2 +\Delta^2}}{\Delta}{\rm sh}
\alpha_n +\frac{\omega_n}{\Delta}{\rm ch}\alpha_n\right ]^2 +\cos^2 (\Phi
{\rm tg}\theta\cos\varphi)},
\end{split}
\label{5}
\end{equation}
where $A =2ek_F^2/\pi^2$, $\omega_n = (2n+1)\pi T$, $2\Delta$ is the
superconductor gap, $\alpha_n = 2\omega_n d/v_F \cos\theta$, and $\Phi$ is given
by Eq. (\ref{3}). The function $j^d (\Phi)$ is noted for interesting features.
In small magnetic fields ($\Phi \ll 1$) $j^d\approx -j_s \Phi$. Such low
fields can lead to the effect of extrascreening of the external magnetic field
(see \cite{24}). When the field increases ($\Phi\simeq 1$), the current starts
oscillating and for certain "phases" it turns to zero at regular intervals
"phases" $\Phi$. With high values of the inequality ($\Phi \gg 1$), the current
amplitude decreases.

\section{Resonance spikes in the density of states of $NS$ structure in weak
magnetic fields}
\*
\par In the region of weak magnetic fields, the density of states of the quasiparticles
that are described by the spectrum of Eq. (\ref{2}) exhibits sharp singularities.
The spectrum of Eq. (\ref{2}) is formed by the trajectories of the quasiparticles
which collide with the dielectric and superconducting boundaries. It encloses a
certain area penetrated by a magnetic flux. At any instant when the magnetic flux becomes
a multiple of the superconducting flux quantum, the density of states experiences
resonance spikes.

Let us consider the cross-section of a $NS$ structure. Assume that the superconducting
cylinder radius $R$ and the normal layer thickness $d$ have a mesoscopic scale. The
density of states $\nu (\varepsilon)$ can be calculated proceeding from the expression
\begin{equation}
\nu(\varepsilon) =\sum_{n,\alpha,\sigma} \int dq \delta[\varepsilon - \varepsilon_n
(q,\alpha)].
\label{6}
\end{equation}

The summation is taken over all quantum numbers $n$, $q$, $\alpha$ and
spin $\sigma$. Since we are not interested in the contribution from the states
formed by the trajectories of the quasiparticles with $\alpha >\alpha_c$, we
can write down
\begin{equation}
\nu(\varepsilon) = \int\limits_{-\alpha_c}^{\alpha_c} d\alpha \nu (
\varepsilon ; \alpha),
\label{7}
\end{equation}
where $\nu (\varepsilon; \alpha)$ is the contribution to the density of states
from the pre-assigned trajectory with a fixed $\alpha$. Eq. (\ref{2}) for the
low-lying Andreev levels ($\varepsilon \ll \Delta$) is taken as a spectrum. After
integration with respect to $q$ and introduction of the notation $\beta =\pi\hbar
/2dm^*$,
we can pass on to the dimensionless energy $\epsilon = \varepsilon/\beta p_F$.
For $\nu (\varepsilon,\alpha)$ we have the expression
\begin{equation}
\nu(\epsilon,\alpha) = \frac{2p_F}{\pi^2\beta d} \epsilon^2
\sum_n \frac{{\rm sec}^2
\alpha\,\,\theta[\vert n+\varkappa\vert -\epsilon {\rm sec}\alpha]}
{(n+\varkappa)^2 \sqrt{(n+\varkappa)^2 -\epsilon^2 {\rm sec}^2 \alpha}},
\label{8}
\end{equation}
where $\varkappa = 1/2 -\Phi {\rm tg} \alpha/\pi$, and
$\theta (x)$ is the stepwise Heaviside function. Eq. (\ref{8}) suggests two cases
depending on the parameter $n+\varkappa$.

a){\underline{Non-resonance case.}} If $n+\varkappa\not= 0$, the energy dependence
under the radical sign in Eq. (\ref{8}) can be neglected for small energies (
$\epsilon\to 0$). Then, the nonresonance contribution to the density of states
is
\begin{equation}
\nu^{(0)} \sim  \frac{2p_F}{\pi^2\beta d} \epsilon^2
\int\limits_{0}^{\alpha_c}
d\alpha \sum_{n=-\infty}^{+\infty} \frac{{\rm sec}^2\alpha}{(n+\varkappa)^3}.
\label{9}
\end{equation}
The series in Eq.(\ref{9}) is calculated readily by the formula in \cite{27}:
$$
\sum_{k=-\infty}^{+\infty} \frac{1}{(k+\varkappa)^n} = (-1)^{n-1} \frac{\pi}{(n-1)!}
\frac{d^{n-1}}{d\varkappa^{n-1}} {\rm ctg} \pi \varkappa.
$$
After calculation of the integral we obtain
\begin{equation}
\nu^{(0)} \sim \frac{p_F}{\beta d} \epsilon^2 \frac{\Phi_0}{a}
{\rm tg}^2 \left [ \frac{2\pi a}
{\Phi_0} \sqrt{\frac{2R}{d}}\right ],
\label{10}
\end{equation}
where $\sqrt{\frac{2R}{d}}\simeq {\rm tg}\alpha_c$.

b){\underline{Resonance case.}} Now we go back to Eq. (\ref{8}). We find
$\nu^{{\rm res}}$ as
\begin{equation}
\nu^{\rm res} \sim \epsilon^2 \int\limits_0^{\alpha_c} d\alpha \sum_n \frac{
{\rm sec}^2\alpha\theta[\vert {\mathfrak a}_n-b{\rm tg}\alpha\vert
-\epsilon
{\rm sec}\alpha]}
{\vert {\mathfrak a}_n - b{\rm tg}\alpha\vert^2 \sqrt{\vert {\mathfrak a}_n -
b{\rm tg}\alpha\vert^2 - \epsilon^2{\rm sec}^2\alpha}},
\label{11}
\end{equation}
where the notations ${\mathfrak a}_n = n+1/2$,
$b=\frac{2a}{\Phi_0}$ are introduced. Eq. (\ref{11}) shows that at certain values
of the flux (b), the radicand in the denominator turns to zero.

Prior to calculation of $\nu^{\rm res}$, let us discuss the question of the contribution
of different angles $\alpha$ to the resonance amplitude. It is reasonable to
assume that because of the factor ${\rm sec}^2 \alpha$ in the numenator of Eq. (\ref{11}),
the angles $\alpha\sim\alpha_c$ are the main contributors to the integral.
It is convenient to employ in the integral a new variable of integration $x={\rm tg}
\alpha$. Then the neighborhood of the upper limit $x_0 = {\rm tg} \alpha_c$ is the
main contributor to the integral. Introducing the notation $\tilde a = {\mathfrak a}_n
- bx_0$ and the small deviation $\xi = x_0 -x\ll 1$, we can write down the equation
for the resonance condition as:
\begin{equation}
(b^2 - \epsilon^2)\xi^2 + 2(\tilde ab +\epsilon^2 x_0)\xi +\tilde a^2 -
\epsilon^2 (1+x_0^2) =0.
\label{12}
\end{equation}
The point of our interest is the asymptotics $\nu (\epsilon)$ at low $\epsilon\to 0$.
Eq. (\ref{12}) is solved to the accuracy within first-order terms of $\vert
\epsilon\vert$:
\begin{equation}
\xi_{1,2} \simeq \frac{\tilde a}{b} \pm \frac{\vert\epsilon\vert}{b}
\sqrt{1+x_0^2}.
\label{13}
\end{equation}

The expression in front of the radical in the denominator of Eq. (\ref{11})
has the second order smallness in $\vert\epsilon\vert$, i.e.
$\vert \tilde a\vert^2
\gtrsim\vert\epsilon\vert^2 (1+x_0^2)$, which leads to its cancellation with
the similar small parameter in the numenator.

The remaining integral is estimated to be a constant of about unity. The resonance
-induced spike of the density of states always appear when the Andreev level
coincides with the Fermi energy at a certain flux in the $N$ layer. In the vicinity
of the chemical potential there is a strong degeneracy of the quasiparticle states
 with respect to the quantum number $q$. As a result, a macroscopic number of $q$
states contribute to the amplitude of the effect. Near the resonance, the ratio
of the resonance and nonresonance amplitudes of the density of states is
\begin{equation}
\frac{\nu^{\rm res}}{\nu^{(0)}} \sim \frac{1}{\vert\epsilon\vert^2}\gg 1.
\label{14}
\end{equation}

It is thus shown that a change in the magnetic flux leads to resonance
spikes in the density of states of the $NS$ structure. The flux interval between the
spikes is equal to the superconducting flux quantum $\Phi_0$.

\section{Calculation of susceptibility of $NS$ contact}
\*
\par To explain the reentrant effect, we need to have an expression for the susceptibility
of the $NS$ structure. We assume that in a weak
magnetic field the total susceptibility of the $NS$
sample consists of two contributions. Firstly, the response of the superconductor
to the applied magnetic field generates the Meissner effect.
Note that the diamagnetic response is observed in all fields up to the
critical one. The amplitude of the diamagnetic current increases monotonously
with lowering temperature.
On the
other hand, the presence of a pure normal metal in the $NS$ structure produces a
paramagnetic contribution. In a weak magnetic field the contribution is due to the
Aharonov--Bohm effect and the quantization of the quasiparticle spectrum of the
mesascopic system. When the penetrability of the barrier between the metals is small,
the electrons of the normal metal are reflected specularly from its boundaries.
As compared to the diamagnetic contribution from the superconductor, the paramagnetic
contribution produced by the $N$ layer has a small amplitude and can therefore
be neglected. Thus, the paramagnetic and diamagnetic contributions cannot
compete in the absence of the proximity effect in the $NS$ structure. However,
if the penetrability of the barrier is close to unity, the mechanism of the
Andreev reflection becomes active at the $NS$ boundary
The quasiparticle spectrum of the $N$ layer undergoes a significant
transformation and resonance spikes appear in the amplitude of the density
of states in a certain regions of magnetic fields and temperatures. Simultaneously,
the distribution of the vector potential field in the normal layer becomes
inhomogeneous. As shown below at certain values of the parameters of the problem, the paramagnetic contribution
to the susceptibility of the $NS$ structure can become equal to the diamagnetic
contribution. This is the reason why the reentrant effect appears in pure
mesoscopic $NS$ structures.

Theoretically, the resulting susceptibility including the reentrant effect
can be represented as a sum of the paramagnetic contribution $\chi^p$ of the
$NS$ structure caused by the Andreev scattering and the diamagnetic
susceptibility $\chi^d$ of the system in which there is no proximity effect between
the $N$ and $S$ metals. The temperature-induced behavior of the diamagnetic
current in such a system is well known. As the temperature decreases, the
diamagnetic current amplitude increases and becomes saturated at temperatures
about several millikelvins. At high temperatures $k_B T\gg \hbar V_F/d$, the
diamagnetic current decreases rapidly following the law $j\sim T^{-1}\exp
(-4\pi k_B Td/\hbar V_F)$. Note that in a $NS$ structure in which the electrons
are reflected specularly at both boundaries of the normal metal, the susceptibility
is negative (i.e. diamagnetic) in the whole interval of temperatures $0<T<T_c$.
However, we will not use this approach to estimate the resulting susceptibility.
Below we calculate the screening current of the $NS$ structure.
It naturally allows for the paramagnetic contribution at certain values of the
magnetic field and temperature. We focus our attention on calculation of the
paramagnetic contribution in structures with a pronounced proximity effect.
This is important especially in the context of the recent statement \cite{28}
that no paramagnetic reentrance can occur in $NS$ proximity cylinders in the
absence of electron-electron interaction in the $N$ layer.

a) {\underline{ Paramagnetic susceptibility of $NS$ contact}}

The contribution of the states in Eq(\ref{2}) to the paramagnetic susceptibility
of the normal layer in a $NS$ contact can be calculated proceeding from
the expression for the thermodynamic potential ($k_B=1$)
\begin{equation}
\Omega =-T\sum_{n,q,\alpha\atop\sigma} \ln[1+\exp(-\varepsilon_n(q,\alpha)/T)],
\label{15}
\end{equation}
where the summation is taken over the spin ($\sigma$) and all the states related to
the trajectories of the quasiparticles with $a\lesssim a_c$. The expression for
susceptibility (per unit volume $V$ of the normal metal) is found using the
formula
$$
\chi=-\frac{1}{V} \frac{\partial^2\Omega}{\partial H^2}.
$$

After performing the summation over the spin and taking into account two
signs of the angle $\alpha$ and of the quasimomentum component $q$, we
arrive at the initial expression for paramagnetic susceptibility ($\zeta$ is
the chemical potential of the metal):
\begin{equation}
\begin{split}
\chi & =  \frac{d}{2T{m^*}^2\Phi_0^2}
\int\limits_{-\zeta}^\infty \frac{d\varepsilon
\exp(\varepsilon/T)}{[\exp(\varepsilon/T)+1]^2} \times\cr
& \times \sum_n\int\limits_0^{\alpha_c}
d\alpha\cos\alpha\sin^2\alpha\int\limits_0^{p_F} dq(p_F^2-q^2)^{3/2}
\delta(\varepsilon - \varepsilon_n(q,\alpha)).
\end{split}
\label{16}
\end{equation}

In Ref.\cite{20} we lost one of the radicals $(p_F^2-q^2)^{1/2}$ in the
similar initial expression for $\chi$. As a result, the amplitude of the
paramagnetic contribution appeared to be underestimated. This mistake is
corrected in this work.

It is convenient to present the spectrum in terms of $\beta=\frac{\pi\hbar}
{2m^* d}$ and $\varkappa=\frac{1}{2}-\frac{{\rm tg}\alpha}{\pi}\Phi$ as
$\varepsilon_n(q,\alpha) = \beta\cos\alpha(n+\varkappa)\sqrt{p_F^2-q^2}$.
Now we introduce the dimensionless energy $\epsilon =\frac{\varepsilon}{\beta
p_F}=\frac{\varepsilon}{\delta\varepsilon}$, $\delta\varepsilon = \frac{\pi
\hbar V_F}{2d}$ is the distance between the Andreev levels in the $SN$
structure. Since $\zeta/\delta\varepsilon \gg 1$, the lower limit of the energy
integral can be replaced with $-\infty$. By introducing the variable $x=
{\rm tg}\alpha$ and the notation ${\mathfrak a}_n = n+1/2$, $b=b(H,T) =
2a/\Phi_0$, $a=\int\limits_0^d A(x) dx$, $x_0 = {\rm tg} \alpha_0 =
\sqrt{2R/d}$ and taking into account the parity of the integrand we obtain,
instead of Eq. (\ref{16}):
\begin{equation}
\chi=C\int\limits_0^\infty\frac{d\epsilon \epsilon^4}{{\rm ch}^2 (\eta\epsilon
/2)}\sum_{n=0}^{n_0}\int\limits_0^{x_0} \frac{dx x^2}{({\mathfrak a}_n -bx)^4}
\frac{\theta[{\mathfrak a}_n -bx-\epsilon\sqrt{1+x^2}]}{\sqrt{({\mathfrak a}_n
-bx)^2-\epsilon^2(1+x^2)}}.
\label{17}
\end{equation}

In Eq.(\ref{17}) the summation is taken over the quantum numbers of the
"particles". Here $C=\frac{\zeta^2d}{T\Phi_0^2}$, $\eta=\frac{\delta\epsilon}
{T}$, $n_0$ is the number of Andreev levels in the potential well and
$\theta$ is the Heaviside step function. It is seen in Eq.(\ref{17}) that for the
given "subzone" $n$ the amplitude of the paramagnetic susceptibility increases sharply
whenever the Andreev level coincides with the chemical potential of the metal.
The resonant spike of susceptibility occurs when ${\mathfrak a}_n -bx$ tends
to zero on a change in the magnetic field (or temperature). Because of the
finite number of Andreev levels, the existence region of the isothermal reentrant
effect is within $0<H\lesssim H_{{\rm max}}$.

Let us calculate the integral over $x$ in Eq.(\ref{17}). It contains a singularity
under the radical $R(x)=\sqrt{Ax^2+Bx+C}$ where $A=b^2-\epsilon^2$, $B=
-2{\mathfrak a}_n b$, $C={\mathfrak a}_n^2 -\epsilon^2$. The singularity is
determined by the roots of the quadratic equation $x_{1,2} =\frac{{\mathfrak a}_n
b}{b^2-\epsilon^2} \pm\frac{\vert\epsilon\vert}{b^2-\epsilon^2} \sqrt{b^2 +
{\mathfrak a}_n^2 -\epsilon^2}$. On introducing the notation $\alpha_0 =
\frac{{\mathfrak a}_n}{b}$, the expression for the roots can be written with
a linear accuracy with respect to $\epsilon$ as
\begin{equation}
x_{1,2}\simeq\alpha_0 \pm\frac{\vert\epsilon\vert}{b}\sqrt{1+\alpha_0^2}.
\label{18}
\end{equation}

The main contribution to the integral over $x$, Eq.(\ref{17}), is made by the
vicinity of the point $\epsilon\to 0$. If we exclude the singular points from the
interval of integration, the indefinite integral over $x$ can be calculated accurately
(see the details in the Appendix). Because the $\theta$-function is present under
the integral, the integration intervals $(0,x_1)$ and $(x_2,x_0)$ make a
finite contribution to the integral. On substituting the limits of integration,
the expressions obtained have different powers of the parameter $\vert\epsilon\vert^{-1}$.
We retain only the most important terms in order $\vert\epsilon\vert^{-4}$
that determine
amplitude of the effect. The discarded terms have higher orders of $\epsilon$
-smallness. The intervals $(0,x_1)$ and $(x_2,x_0)$ make contributions of the same
order of $\epsilon$-magnitude. The region $(x_1,x_2)$ does not contribute to the
integral at all.

The estimate for the integral over $x$ is
\begin{equation}
\frac{4}{3}\frac{\alpha_0^2}{b(1+\alpha_0^2)^2}\frac{1}{\epsilon^4}.
\label{19}
\end{equation}
On substituting Eq.(\ref{19}) into Eq.(\ref{17}), the parameter $\epsilon^4$
drops out of the energy integral and we can take it quite easily. Taking
into account the energy limits $\theta({\mathfrak a}_n -\vert\epsilon\vert)$
appearing in the process of calculation we can obtain the expression for the
paramagnetic contribution to the susceptibility of the $NS$ structure, which in
dimensional units has the form
\begin{equation}
\chi^p\simeq\frac{16\zeta^2d^2}{3\pi\hbar V_F\Phi_0^2}\sum_{n=0}^{n_0}
\frac{b(H,T) {\rm th}\left[ \frac{\pi\hbar V_F}{4dk_BT}(n+1/2)\right]}
{(n+1/2)^2 \left [1+\left (\frac{b(H,T)}{n+1/2}\right )^2\right ]^2}.
\label{20}
\end{equation}

In Eq.(\ref{20}) the summation over the quantum number $n$ is taken within
finite limits, where $n_0$ has the meaning of the maximum number of the Andreev
levels inside the potential well of the $NS$ structure. Its order of magnitude
in $n_0\sim \Delta/\delta\varepsilon$, where $\delta\varepsilon$ is the
distance between the Andreev levels, $\delta\varepsilon = \pi\hbar V_F/2d$,
and $2\Delta$ is the energy gap. The flux $b(H,T)=2a/\Phi_0$ depends on
both the magnetic field and temperature. In the pre-assigned field its
value is dictated by the screening current of the $NS$ structure $j=-j_s
\varphi (a/\Phi_0)$ (see Eq.(\ref{4})). The obtained expression for $\chi^p$
manifests a more rapid decrease susceptibility at the increasing parameter
$b(H,T)$ than it was evidenced by Eq.(\ref{5}) in Ref.\cite{20}.

We first discuss the isothermal case of a very low temperature and clear up the
qualitative behavior of susceptibility in Eq.(\ref{20}). We shall proceed from
the region of very strong magnetic fields ($a/\Phi_0 \gg 1$) in which the second
term in Eq.(\ref{4}) is negligible. Then the dimensionless flux $b(H,T)\gg 1$
and the amplitude of the paramagnetic contribution in Eq. (\ref{20}) decreases
as $b(H,T)$ raised to power $3$. In comparatively weak magnetic fields $a/
\Phi_0 \sim 1$), the function $\varphi (x)$ is actually an oscillating function of
$H$ and here we can expect the reentrant effect. Indeed as the field decreases
to a certain value and the parameter $b(H,T)/n_0$ becomes $\sim 1$ ($n_0$
is the number of the Andreev levels in the potential well), the amplitude of the
paramagnetic susceptibility of the $NS$ structure accepts for the first time an
appreciable contribution from the highest Andreev "subband" (level). On a
further decrease in this field, the contribution from the highest "subband"
persists, but in a certain lower field an additional contribution appears from
the neighboring lower-lying "subband" $n_0 -1$. Finally, in a very weak field all
the "subbands" of the $NS$ structure start to contribute and the paramagnetic
susceptibility amplitude reaches its peak. However, at $H\to 0$ ($a/\Phi_0 \to 0$),
the paramagnetic contribution turns to zero, as follows from Eq. (\ref{20}).
The reason is that the resonance condition for the Andreev levels (Eq.(\ref{2})),
cannot be realized in a zero field.

Now we change to the case when the temperature of the $NS$ structure varies but
the field is kept constant. We assume the field to be weak ($H\sim 2\cdot 10^{-1}
Oe$). The second term in Eq.(\ref{4}) for the flux is very important.
It is highest at millikelvin temperatures. As a result, the parameter $b(H,T)$
has the lowest value. In this temperature region the hyperbolic tangent is close to
unity and the paramagnetic contribution is dependent only on the parameter
$b(H,T)$. Under this condition, all the "subbands" of the $NS$ structure
contribute to the amplitude of the effect. As the temperature rises, the parameter
$b(H,T)$ increases smoothly. Simultaneously, the argument of the hyperbolic
tangent decreases. At a certain temperature, when the condition $k_B T> \pi
\hbar V_F /4d$ is met, the contribution from the lowest "subband" starts dying
down and its amplitude is decreasing linearly with growing $T$. On a
further rise of the temperature, the contributions from the higher "subbands" of
the spectrum die down in succession. Finally, at a very high temperature the
paramagnetic contribution tends to zero.

Let us estimate the amplitude of the paramagnetic contribution. The parameter
$b(H,T)$ is dependent on the value of the flux $a=\int\limits_0^d A(x) dx$,
which at constant $T$ can be found by solving the self-consistent equation
Eq. (\ref{4}). In the region of millikelvin temperatures and magnetic fields
$H\sim 2\cdot 10^{-1} Oe$ the paramagnetic contribution has the largest amplitude.
We obtain $b(H,T)\sim 10^{-4}$ in this region of $T$ and $H$. The coefficient
before the sum in Eq. (\ref{20}) can be found by substituting $\zeta^{Ag}
\simeq 8.75 \cdot 10^{-12} erg$, $d=3.3\cdot 10^{-4} cm$ $V_F^{Ag}\sim
1.39\cdot 10^8 cm/sec$ for the characteristic parameters of the normal $Ag$
layer. We thus obtain $16\zeta^2 d^2/3\pi\hbar V_F \Phi_0^2 \simeq 2.418\cdot
10^3$. The product of this coefficient and the parameter $b(H,T)$ yields the
order of magnitude of the paramagnetic contribution amplitude. It is seen that
the largest amplitude of the paramagnetic contribution exceeds that of the
diamagnetic contribution in the vicinity of $T=0$.

b){\underline{Full magnetic susceptibility of $NS$ structure in the presence of}}\\
{\underline{proximity effect}}

Let us consider a structure in which the electrons experience the Andreev
scattering at the $NS$ boundary. In the presence of magnetic field, the
screening current is induced in the normal layer due to the Meissner effect.
We estimate the susceptibility generated by this current.

The total current $J$ is related to the magnetic moment $M$ as
\begin{equation}
M=\frac{1}{c} J S_0,
\label{21}
\end{equation}
where $S_0 \simeq \pi R^2$ is the cylinder cross-section ($d\ll R$). Let the
average current density be $j$. The total current is then $J=Sj$, where
$S=dL$ ($L$ is the cylinder generatrix). The density of the screening current in
$NS$ proximity sandwiches was calculated by Zaikin \cite{24},\cite{28}.
We reproduce the formula for the current density (see Eq. (\ref{5})), which is
valid at arbitrary values of temperature and magnetic field. At $T\ll \hbar V_F
/d$ it is
\begin{equation}
j(\Phi) \simeq -\frac{4ek_F^2 T}{\pi^2} \sum_{\omega_n >0} \int\limits_0^{\pi/2}
d\theta \int\limits_0^{\pi/2} d\varphi \sin^2\theta\cos\varphi
\frac{\sin[2{\rm tg}\theta\cos\varphi \Phi]}{{\rm sh}^2\alpha_n +\cos^2
[{\rm tg}\theta\cos\varphi\Phi]}.
\label{22}
\end{equation}
Here $\alpha_n=\frac{2\omega_n d}{V_F\cos\theta}$, $\omega_n =(2n+1)\pi T$ and the
phase $\Phi$ follows from Eq.(\ref{3}). Near $T=0$ the summation of frequencies
in Eq. (\ref{22}) can be replaced with integration. For $\Phi \le 1$ the response
of the current is \cite{28}
\begin{equation}
j(\Phi)\simeq -\frac{ek_F^2 V_F}{\pi^3 d} \int\limits_0^{\pi /2}d\theta
\int\limits_0^{\pi /2} d\varphi \sin^2\theta\cos\theta\cos\varphi \sin[2\Phi
{\rm tg} \theta\cos\varphi].
\label{23}
\end{equation}
If the field is small enough to meet the condition $\Phi\ll 1$, Eq. (\ref{23})
reduces to the result that was obtained for the first time in \cite{24}:
\begin{equation}
j(\Phi) = -\frac{ek_F^2 V_F}{6\pi^2d}\Phi.
\label{24}
\end{equation}
At "phases" $\Phi\gg 1$, the screening current of Eq. (\ref{23}) turns to zero.

The current-phase dependence at $T=0$ is plotted in Fig.2. The dependence is
nonlinear and its amplitude has a maximum at a certain value of $\Phi$. Knowing
the current-phase dependence, we can determine the susceptibility of the $NS$
structure using the equation $\chi = dM/dH$. It is seen in Fig.2 that the
susceptibility $\chi$ of the $NS$ structure (the derivative of current with
respect to field) changes its sign at a certain low value of the magnetic field
$H_r$. The susceptibility is "diamagnetic" in the region of high magnetic fields
and "paramagnetic" at $H<H_r$. The "paramagnetic" portion of the curve is due to the
proximity effect at the $NS$ boundary and to the Andreev levels in the $N$ layer.

\begin{figure}
\centering
\includegraphics[height=8cm,width=5cm]{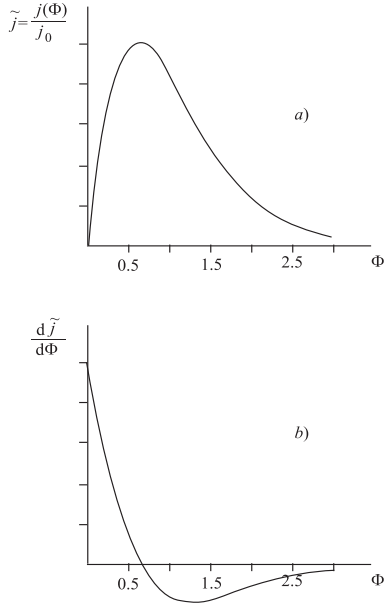}
\caption{a) Dependence of screening current Eq. (5) on "phase" $\Phi$ at 
$T=0,092 K$. Current is in arbitrary units. b) The derivative of current with
respect to "phase" (magnetic field) changes its sign in the region of low
$\Phi$-values. When the magnetic field tends to zero the full magnetic susceptibility
of $NS$ structure is positive.}
\end{figure}

Let us estimate $\chi$ in the linear-response regime near $T=0$ when this dependence
is described by Eq.(\ref{24}). In such weak fields we obtain $\Phi\simeq \frac
{3\pi H\lambda_N^2 (T)}{\Phi_0}$, where the "penetration depth" $\lambda_N$
into  the normal metal is dependent on temperature \cite{26}:
\begin{equation}
\begin{split}
\lambda_N^{-2} (0) = \frac{4\pi ne^2}{m^* c^2}; & \quad(T\simeq 0)\\
\lambda_N^{-2}\sim\lambda_N^{-2} (0) \frac{6T}{T_A} \exp(-2T/T_A); &
\quad (T\gg T_A=\frac{\hbar V_F}{2\pi d}).
\end{split}
\label{25}
\end{equation}

The estimate of susceptibility in the millikelvin region is $\chi\sim-
\frac{R}{4c}\frac{ek_F^2 V_F}{\pi d} \frac{\lambda_N^2 (0)}{\Phi_0}$. For
the parameters of the problem $d=3.3\cdot 10^{-4} cm$, $R=8.2\cdot 10^{-4} cm$,
$k_F^{Ag} \sim 1.2\cdot 10^8 cm^{-1}$, $V_F^{Ag} \sim 1.39\cdot 10^8 cm/sec$,
$\lambda_N (0) \sim 2\cdot 10^{-6} cm$ we obtain $\chi=-0.06$, which is close to $\chi
=-\frac{3}{4}\frac{1}{4\pi}$.

Now we keep the magnetic field (assuming it weak) constant and plot the screening
current versus temperature in a wide $T$-range. This dependence plotted
using Eq. (\ref{5}) is shown in Fig.3. It is seen that the current amplitude has
a maximum at a certain $T_r$.

\section{Discussion}
\*

In this study we have investigated the behavior of a superconducting cylinder
covered with a thin layer of a pure normal metal. It is assumed that the normal
metal and superconductor are in good contact. The system was placed in a
magnetic field directed along the $NS$ boundary. The $NS$ structure has mesoscopic
scale dimensions. It is assumed that the mean free path of the quasiparticles in the
$N$ layer exceeds the characteristic length $\xi_N = \hbar V_F/k_BT$, which has the
meaning of the coherence length for a system with disturbed long-range order.
The goal of this study was to interpret the experiments in which A.C.Mota et
al. \cite{10} -- \cite{14} observed anomalous behavior of the magnetic 
susceptibility of a $NS$ structure at varying temperature in a constant 
magnetic field or in a varying magnetic field
at a constant temperature. This
phenomenon was called a reentrant effect. Until recently it has not been explained
adequately.

Earlier \cite{20} the author clarified the nature of the reentrant effect. As was
found, the origin of the paramagnetic contribution is closely connected with
the properties of the quantized Andreev levels that are dependent on the magnetic
flux varying with both temperature and magnetic field. Typically, the levels in
the $NS$ structure time from time (at certain values of the field $H$ or
temperatures) coincide with the chemical potential of the metal. As a result,
the state of the system is highly degenerate and the density of states of the $NS$
structure experiences resonance spikes. The response of the normal mesoscopic
layer to a weak magnetic field is paramagnetic.

A theory of the reentrant effect has been developed in this study. We calculated
the paramagnetic contribution separately and analyzed its behavior in a varying
magnetic field and at varying temperature. In the course of this calculation we
corrected the mistake made in \cite{20} which led to underestimation of the
effect amplitude. The paramagnetic response is determined only by the trajectories
of the quasiparticles that collide with the $NS$ boundary. It is shown that the
reentrant effect can occur in a certain range of weak magnetic fields at
temperatures no higher than $100 mK$. We believe that paramagnetic reentrant
effect is an intrinsic effect of mesoscopic $NS$ proximity structures in the
very low temperature limit.

Assume that the temperature of the
$NS$ structure is about $10^{-3} K$ and the magnetic field is increasing. As soon
as the field exceeds a certain value $H_r$, the isothermal  reentrant effect must
vanish. In strong fields the Andreev levels cease to make a resonance contribution
to the paramagnetic susceptibility. Now the paramagnetic contribution is made by
the states formed by the trajectories of the quasiparticles that collide only with
the dielectric boundary. However, their contribution to the resulting susceptibility
of the structure is small because of the smallness of the quasiclassical parameter
of the problem $1/k_F R$. Under this condition the susceptibility exhibits
diamagnetic behavior in all strong fields up to the critical one.

A self-consistent calculation of the screening current of the $NS$ structure was
performed taking into account the contribution from the Andreev levels. The analysis
of the derived expression suggests the paramagnetic contribution to current.
For example, Fig.2 illustrates the dependence of the current upon the phase
(magnetic field). The values of the current $j$ to the left of the extremum
$\Phi_r$ account for the contribution of the Andreev levels. The derivative of
this curve with respect to the field is proportional to the magnetic susceptibility
of the $NS$ structure. It is positive ("paramagnetic") in the region of low magnetic
fields and negative ("diamagnetic") in high fields.

Similar behavior is observed when the susceptibility of the $NS$ structure is
measured as a function of temperature in a pre-assigned weak magnetic field: it
is "paramagnetic" in the region $T<T_r$ and "diamagnetic" at $T>T_R$ up to the
critical temperature. Temperature dependence of magnetic susceptibility in
the $NS$ structure at fixed magnetic field will be investigated in detail in
separate publication.

In the absence of the proximity effect in the $NS$ structure, when the penetrability
of the barrier between the $S$ and $N$ metals is small, the electrons of the normal
metal are reflected specularly from its boundaries. In this case the $SN$ structure
is a total of two isolated subsystems (normal metal and superconductor) placed
into a magnetic field. Because of the Meissner effect, diamagnetic current develops
near the superconductor surface. In normal metal, because of the Aharonov-Bohm
effect, the quantized spectrum of quasiparticles is dependent on the magnetic
flux through the cross-section of the cylinder. The flux generates a paramagnetic
contribution to the susceptibility whose quasiclassical parameter of the problem
$1/k_F R$ is small. Hence, in the absence of the proximity effect no competition
is possible between the paramagnetic and diamagnetic contributions in the $NS$
structure, and the reentrant effect is unobservable in such $NS$ sample.

To conclude, it should be noted that the explanation proposed in this study
for the reentrant effect was developed within a model which does not assume the
electron-electron interaction in the $N$ layer of the $NS$ structure. In terms
of the free-electron model, a large paramagnetic contribution to the susceptibility
of the $NS$ structure appears in the region of very low temperatures in a weak
magnetic field. If we increase the thickness $d$ of the pre-assigned normal metal,
this would lead to a greater number of the Andreev levels $n_0$ in the potential
well and affect the solutions of the self-consistent equation for $a$.
As a result, the shape of the curve of the paramagnetic susceptibility would be
slightly "deformed" though its qualitative behavior would remain the same.
\newpage

{\bf Appendix}
\setcounter{equation}{0}
\renewcommand{\theequation}{A\arabic{equation}}

Let us calculate the integral taken over $x$ in Eq. (\ref{15}):
\begin{equation}
J=\int\limits_0^{x_0} \frac{dx x^2 \theta[{\mathfrak a}_n - bx - \epsilon
\sqrt{1+x^2}]}{({\mathfrak a}_n - bx)^4 \sqrt{({\mathfrak a}_n - bx)^2 -
\epsilon^2(1+x^2)}}.
\label{A1}
\end{equation}

After introducing the notation $\alpha_0 = {\mathfrak a}_n/b$, we can see that
the function in front of the radical in the denominator has a singularity at the
point $x=\alpha_0$. Besides, as was noted in the text, the integrand has singularities
at the points $x_1$, $x_2$.

Integral (\ref{A1}) can be written as a sum of four integrals
$$
J=\int\limits_0^{x_0} dx\ldots = \lim\limits_{\varepsilon\to 0}
\left \{  \int\limits_0^{x_1-\varepsilon} dx\ldots +
\int\limits_{x_1+\varepsilon}^{\alpha_0-\varepsilon} dx\ldots +
\int\limits_{\alpha_0+\varepsilon}^{x_2-\varepsilon} dx\ldots +
\int\limits_{x_2+\varepsilon}^{x_0} dx\ldots\right \}.
$$

It is obvious that the presence of the $\theta$-function makes the second and
the third integrals equal to zero. We first calculate the integral $J_1$:
\begin{equation}
J_1 = \frac{1}{b^4} \lim\limits_{\varepsilon\to 0} \int\limits_0^{x_1-\varepsilon}
\frac{dx x^2}{(\alpha_0 - x)^4 \sqrt{Ax^2 +Bx +C}},
\label{A2}
\end{equation}
where $A=b^2 -\epsilon^2$, $B=-2{\mathfrak a}_n b$, $C= {\mathfrak a}_n^2 -
\epsilon^2$. On substituting the variable $\alpha_0 -x = 1/t$, the indefinite integral
becomes $\int\frac{dt t(\alpha_0 t-1)^2}{\sqrt{\alpha t^2 +\beta t +A}}$, where
$\alpha =-(1+\alpha_0^2)\epsilon^2$, $\beta=2\alpha_0\epsilon^2$. It can be calculated
by the method of undetermined coefficients:
\begin{equation*}
\begin{split}
\int\frac{dt f(t)}{\sqrt{\alpha t^2+\beta t +\gamma}} & = ( A_1 t^{n-1} +
A_2 t^{n-2} +\ldots+A_n)\sqrt{\alpha t^2 +\beta t +\gamma} +\\
&+ A_{n+1} \int\frac{dt}{\sqrt{\alpha t^2 +\beta t +\gamma}}
\end{split}
\end{equation*}
if $f(t)$ is the polinomial to power $n$. Although the calculation is tedious, it
is actually simple. The coefficients $A_1$, $A_2$, $A_3$ and $A_4$ are readily found
as:
$$
A_1 =- \frac{\alpha_0^2}{3\epsilon^2 (1+\alpha_0^2)}, \quad
A_2 = \frac{\alpha_0(1+\alpha_0^2/\epsilon)}{\epsilon^2(1+\alpha_0^2)^2},
$$
$$
A_3 =-\frac{2{\mathfrak a}_n^2}{3\epsilon^4 (1+\alpha_0^2)^2} +
\frac{-3+5\alpha_0^2+\alpha_0^4/2}{3\epsilon^2 (1+\alpha_0^2)^3},
$$
$$
A_4 = \frac{{\mathfrak a}_n^2 (\alpha_0^2/2 -1)}{\epsilon^2 \alpha_0
(1+\alpha_0^2)^2} + \frac{\alpha_0(2-\alpha_0^2/2)}{(1+\alpha_0^2)^3}.
$$
It is seen that the coefficients have different orders of $\epsilon^{-1}$-magnitude:
$A_1, A_2, A_4 \simeq \epsilon^{-2}$, $A_3 \simeq \epsilon^{-4}$. Finally,
we have to calculate six integrals
\begin{equation}
\begin{split}
J_1 &= \frac{1}{b^4}\lim\limits_{\varepsilon\to 0}
\int\limits_{t_0}^{t_1-\varepsilon}dt\bigg\{ 2A_1t\sqrt{R(t)}
+A_2\sqrt{R(t)} +A_1 \alpha t^3/\sqrt{R(t)} +\\
&+(A_2\alpha +A_1\beta/2)
\frac{t^2}{\sqrt{R(t)}}+\left (A_3\alpha +A_2\frac{\beta}{2}\right )\frac{t}{\sqrt{R(t)}}
+\frac{(A_3 \beta/2 +A_4)}{\sqrt{R(t)}}\bigg \},
\end{split}
\label{A3}
\end{equation}
where $R(t)=\alpha t^2 +\beta t +A$ and the designations $t_0 =\frac{1}{\alpha_0}$,
$t_1 -\varepsilon = (\alpha_0 -x_1 +\varepsilon)^{-1}$
are introduced. All the six indefinite integrals in expression (\ref{A3}) can be
calculated accurately \cite{29}. After substituting the limits of integration, integrals
$1$, $2$, $3$, $4$ and $5$ are bounded above on energy, which is due to the term
$\sqrt{R(t)} \simeq \sqrt{{\mathfrak a}_n^2 -\epsilon^2}/\alpha_0$, i.e.
$\theta({\mathfrak a}_n -\epsilon)$. Taking into account the determined coefficients
$A_i$ ($i=1,2,3,4$), we can obtain the final expression for $J_1$:
\begin{equation}
\begin{split}
J_1 & \simeq \frac{1}{b^4} \bigg \{ \frac{b(\alpha_0^2 +1/3)}{3\epsilon^2(1+\alpha_0^2)^2}
-\frac{b(1+\alpha_0^2/6)}{\epsilon^2(1+\alpha_0^2)^2} +\frac{2{\mathfrak a}_n^2 b}
{3\epsilon^4(1+\alpha_0^2)^2}+\\
&+\frac{b(1-\frac{5}{3}\alpha_0^2 -\alpha_0^4/6)}{\epsilon^2(1+\alpha_0^2)^3}+
\frac{\alpha_0 b^2(\alpha_0^2/2 -1)}{\epsilon^3(1+\alpha_0^2)^{5/2}}+
\frac{\alpha_0(2-\alpha_0^2/2)}{\epsilon(1+\alpha_0^2)^{7/2}}-\\
&-\frac{b(\alpha_0^2/2-1)}{\epsilon^2(1+\alpha_0^2)^2} -\frac{(2-\alpha_0^2/2)}
{(1+\alpha_0^2)^3 b}\bigg\}.
\end{split}
\label{A4}
\end{equation}
Of all the terms in (\ref{A4}), the most significant contribution is made
by the third term because there is a factor $e^4$ in the numerator of the integral
over the energy in Eq. (\ref{17}). The contributions of the other terms are negligible.
We thus obtain the estimate
\begin{equation}
J_1\simeq \frac{2\alpha_0^2}{3b(1+\alpha_0^2)^2}\frac{1}{\epsilon^4}.
\label{A5}
\end{equation}
A similar calculation of the integral
$$
J_4 = \frac{1}{b^4} \lim\limits_{\varepsilon\to 0}
\int\limits_{x_2+\varepsilon}^{x_0} \frac{dx x^2}{(x-\alpha_0)^4\sqrt{Ax^2 +
Bx+C}}
$$
gives a contribution, which is identical in the order of magnitude with (\ref{A5}).
As a result, we obtain the $J$ estimate present in the text, Eq. (\ref{19}).

The author is sincerely grateful to A.N.Omelyanchouk for helpful discussions
and support, to S.I.Shevchenko for valuable comments.

\end{large}

\begin{thebibliography}{99}

\bibitem{1} Y.Imry, In:Direction in Condenced Matter Physics. B. Grinstein and
G.Mazenko (eds), World Scientific, Singapure (1986), p.101.

\bibitem{2} S.Washburn and R.A.Webb, Adv. Phys. {\bf 35}, 375 (1986).

\bibitem{3} A.G.Aronov and Yu. V.Sharvin, Rev.Mod.Phys. {\bf 59}, 755 (1987).

\bibitem{4} I.O.Kulik,JETP Lett. {\bf 11}, 275 (1970).

\bibitem{5}Y.Aharonov and D.Bohm, Phys. Rev, {\bf 115}, 485 (1959).

\bibitem{6}E.N.Bogachek and G.A.Gogadze, Zh. Eksp. Teor. Fiz, {\bf 63}, 1839 (1972)
[Sov. Phys. JETP, {\bf 36}, 973 (1973)].

\bibitem{7}N.B.Brandt, V.D.Gitsu, A.A.Nikolaeva, and Ya.G.Ponomarev, JETP Lett., {\bf 24},
272 (1976); Zh. Eksp. Teor. Fiz, {\bf 72}, 2332 (1977) [Sov. Phys. JETP {\bf 45}, 1226 (1977)].

\bibitem{8}N.B.Brandt, E.N.Bogachek, D.V.Gitsu, G.A.Gogadze, I.O.Kulik,
A.A.Nikolaeva, and Ya.G.Ponomarev, Fiz.Nizk.Temp. {\bf 8},718 (1982) [Sov. J. Low
Temp. Phys. {\bf 8}, 358 (1982)].

\bibitem{9} A.C.Mota, P.Visani, and A.Pollini. J. Low Temp. Phys. {\bf 76}, 465 (1989).

\bibitem{10} P.Visani, A.C.Mota and A.Pollini, Phys. Rev. Lett., {\bf 65}, 1514 (1990).

\bibitem{11}A.C.Mota, P.Visani, A.Pollini and K.Aupke, Physica {\bf B 197}, 95 (1994).

\bibitem{12} F.B.Muller-Allinger and A.C.Mota, Phys. Rev. Lett.,{\bf 84}, 3161 (2000).

\bibitem{13}F.B.Muller-Allinger and A.C.Mota,Cond-mat/0007331 (2000).

\bibitem{14}R.Frassanito, P.Visani, M.Niderost, A.C.Mota, P.Smeibidl,
K.Swieca, W.Wendler, and F.Pobell, Proceedings LT-21, Part S4-LT; Properties of
Solids 1, Prague (1996), p.2317.

\bibitem{15} A.F.Andreev, Zh. Eksp. Teor. Fiz. {\bf 46} (1964), 1823 
[Sov. Phys. JETP {\bf 9} (1964), 1228]

\bibitem{16}G.A.Gogadze, R.I.Shekhter and M.Jonson, Low Temp. Phys, {\bf 27} (2001),
913 [Fiz. Nizk. Temp. {\bf 27} (2001), 1237]

\bibitem{17}C. Bruder and Y.Imry, Phys. Rev. Lett., {\bf 80} (1998), 5782.

\bibitem{18}A.L.Fauchere, W. Belzig and G.Blatter, Phys. Rev, Lett., {\bf 82} (1999),
3336.

\bibitem{19} K.Maki and S.Haas, Cond-mat/0003413 (2000).

\bibitem{20}G.A.Gogadze, J. Low Temp. Phys., {\bf 31} (2005), 94 [Fiz. Nizk. Temp., {\bf 31}
(2005), 120.

\bibitem{21}G.A.Gogadze, Fiz. Nizk. Temp., {\bf 9} (1983), 1051 [Sov. J. Low Temp.
Phys., {\bf 9} (1983), 543]

\bibitem{22} J.B.Keller and S.I.Rubinow, Ann. Phys. (N.Y), {\bf 9} (1960), 24.

\bibitem{23}I.O.Kulik, Zh. Eksp. Teor. Fiz., {\bf 57} (1969), 1745 [Sov. Phys.
JETP, {\bf 30} (1969), 944].

\bibitem{24} A.D.Zaikin, Solid State Commun., {\bf 41} (1982), 533.

\bibitem{25}W.Belzig, G. Bruder and G.Schon, Phys. Rev. {\bf B53} (1996), 5727.

\bibitem{26} A.L.Faucher and G. Blatter, Phys. Rev. {\bf B56} (1997), 14102.

\bibitem{27} A.P. Prudnikov, Yu. A. Brychkov, and O.I.Marichev, Integrals and
Series (Nauka, Moscow, 1984) (in Russian).

\bibitem{28} A.V.Galaktionov and A.D.Zaikin, Phys.Rev. {\bf B67} (2003), 184518.

\bibitem{29}I.S.Gradshtein and I.M.Ryzhik. Tables of Integrals, Sums, Series,
and Products, Nauka, Moscow (1971).

\end{thebibliography}
\end{document}